\documentclass[reprint, aip,apl,superscriptaddress,showpacs,preprintnumbers,amsmath,amssymb]{revtex4-1}

\usepackage{graphicx}
\usepackage{dcolumn}
\usepackage{bm}
\usepackage{longtable}
\usepackage{hyperref}
\usepackage{lipsum}
\usepackage{braket}
\usepackage{physics}
\usepackage{setspace}
\usepackage{color}
\usepackage{fancyhdr}
\usepackage{titlesec}
\usepackage{float}
\usepackage{upgreek}
\usepackage{array}
\usepackage{epstopdf}

\usepackage[utf8]{inputenc}

\usepackage{wrapfig}
\usepackage{amsmath,amsthm,amssymb}
\usepackage[table,xcdraw]{xcolor}

\begin{document}
\preprint{APS/123-QED}
\title{Geometrically enhanced closed-loop multi-turn sensor devices that enable reliable magnetic domain wall motion}

\author{B. Borie}%
\affiliation{Institut f\"ur Physik, Johannes Gutenberg-Universit\"at Mainz, Staudinger Weg 7, 55128 Mainz, Germany}%
\affiliation{Sensitec GmbH, Hechtsheimer Str. 2, Mainz D-55131, Germany}%

\author{J. Wahrhusen}%
\affiliation{Sensitec GmbH, Hechtsheimer Str. 2, Mainz D-55131, Germany}%

\author{H. Grimm}%
\affiliation{Sensitec GmbH, Hechtsheimer Str. 2, Mainz D-55131, Germany}%

\author{M. Kl\"aui}%
\affiliation{Institut f\"ur Physik, Johannes Gutenberg-Universit\"at Mainz, Staudinger Weg 7, 55128 Mainz, Germany}%

\date{\today}

\begin{abstract}

We experimentally realize a sophisticated structure geometry for reliable magnetic domain wall-based multi-turn-counting sensor devices, which we term closed-loop devices that can sense millions of turns. The concept relies on the reliable propagation of domain walls through a cross-shaped intersection of magnetic conduits, to allow the intertwining of loops of the sensor device. As a key step to reach the necessary reliability of the operation, we develop a combination of tilted wires called the syphon structure at the entrances of the cross. We measure the control and reliability of the domain wall propagation individually for cross-shaped intersections, the syphon geometries and finally combinations of the two for various field configurations (strengths and angles). The various measured syphon geometries yield a dependence of the domain wall propagation on the shape that we explain by the effectively acting transverse and longitudinal external applied magnetic fields. The combination of both elements yields a behaviour that cannot be explained by a simple superposition of the individual different maximum field operation values. We identify as an additional process the nucleation of domain walls in the cross, which then allows us to fully gauge the operational parameters. Finally, we demonstrate that by tuning the central dimensions of the cross and choosing the optimum angle for the syphon structure reliable sensor operation is achieved, which paves the way for disruptive multi-turn sensor devices.

\end{abstract}

\maketitle

The field of magnetic domain walls has generated significant interest since the mid 1960s\cite{Spa66, Spa70, McM97, MK08, All02, All05, Kle08, Oma14, Par08, Don10, Die04, Die07, Die09, Mat12}. The propagation of these magnetic quasi-particles in various magnetic geometries \cite{McM97, MK08} was considered for use in magnetic field sensors. More specifically, the DW can be the active element enabling nonvolatile multiturn-counting sensors \cite{Die04, Die07, Die09, Mat12}. 

An innovative approach was recently proposed based on the simultaneous measurement of several coprime-counting intersected closed-loop architectures \cite{Borsyphon17}. In contrast to the already studied open-loop DW based device structure \cite{Bor17, Mat14}, this alternative concept includes a different geometrical feature, namely a cross-shaped intersection of nanowires, which has been investigated numerically and experimentally \cite{Lew09, Pus13, Bur14, Set16, Borsyphon17}. This geometry ultimately enables a disruptive device that can count millions of turns. Despite the concept having been introduced theoretically and having been studied by micromagnetic simulations \cite{Borsyphon17}, no experimental realization has been reported to date.

In this letter, we identify the externally applied field configurations (strengths and angles) that results in a pinning, propagation, or an unwanted splitting of the DW in the center of cross geometries that leads to a failure event for the sensor. We develop a syphon structure and measure the field configurations that allow for a propagation through the syphon or the pinning in its arm for different syphon tilting angles. We then virtually merge the individual behaviours for DW propagation in a syphon and a cross thus generating the expected behavior of the complete device. We compare the latter to selected real device geometries measured under an applied rotating field thus yielding characteristics for the field regimes that result in the operation. By comparing the complete device and the individual constituents, we identify geometrical parameters that govern the device performance, which allows us to build an optimized sensor with a large field window for reliable operation.

\begin{figure}[ht]
\centering
\includegraphics[scale=0.4]{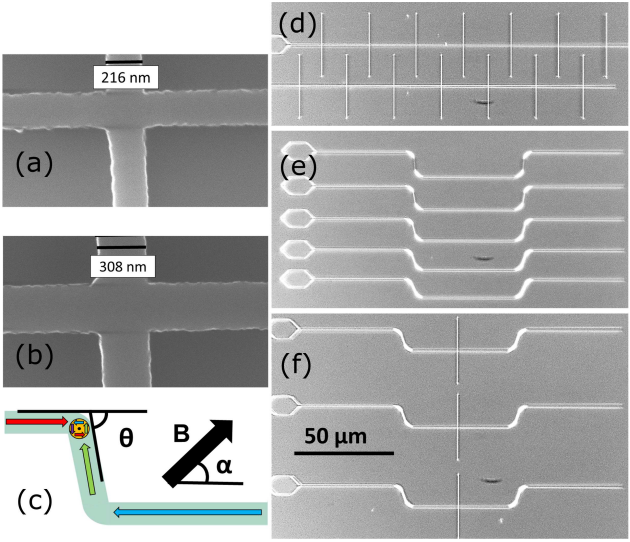}
\caption{ (a) SEM micrograph of a cross with center diagonal length equal to 210$\sqrt{2}$ nm. (b) SEM micrograph of a cross with center diagonal length equal to 300$\sqrt{2}$ nm. (c) Schematic of the syphon initial magnetization configuration. The orange disk with the arrows represents a vortex DW. The colored arrows represent the magnetization direction. The black arrow indicates the applied field with strength B and angle $\alpha$.(d) Optical image of the cross structures with a center diagonal of 300$\sqrt{2}$ nm for the top one and 210$\sqrt{2}$ nm for the bottom structure. (e) Optical microscopy image of the syphon structures. The syphon angles $\theta$ (see (c)) are varied from top to bottom as $\theta$ = 85$^\circ$, 80$^\circ$, 75$^\circ$, 70$^\circ$, and 65$^\circ$. (f) Optical microscopy image of the complete structures with a syphon of angle $\theta$  = 70$^\circ$ and a cross with center diagonal equal to 300$\sqrt{2}$ nm.}
\label{fig1}
\end{figure}

For the fabrication of the structures, a stack of Ni$_{81}$Fe$_{19}$ (30 nm)/ Ta (4 nm) (bottom to top) is deposited on a substrate of SiO$_x$ in a magnetron sputtering. The samples are then patterned by Electron Beam lithography and Ar ion etching.
Scanning electron microscopy (SEM) reveals well-defined structures (Fig. \ref{fig1}). The polycrystalline material used (Ni$_{81}$Fe$_{19}$) is magnetically soft and exhibits a full film coercivity of 2 Oe with a saturation magnetization value of M$_s$ = 795 kA/m. 

The samples are patterned into cross-shaped element (Fig. \ref{fig1} (a, b, and d)), the syphon elements (Fig. \ref{fig1} (c and e)) and in a combination of the two (Fig. \ref{fig1} (f)). A nucleation pad is located at one end of all geometries to allow for the introduction of DWs. Furthermore, wire ends are tapered to avoid DW nucleation at that point.
In the designs, the nominal wire width is 300 nm for all structures as the nucleation and depinning fields for this wire width yield values that are useful for applications \cite{Bor17}. In the center of the cross, the diagonal can be reduced down to 180$\sqrt{2}$ nm.
All experiments are conducted using a magneto-optical Kerr effect microscope (Evico Magnetics). Furthermore, rotating the plane of incidence of the light by 90$^\circ$ enables the observation of the magnetic switching of the structures in orthogonal directions. A vector magnet is utilized for the application of a rotating field up to 100 mT. 

\begin{figure}[ht]
\centering
\includegraphics[scale=0.4]{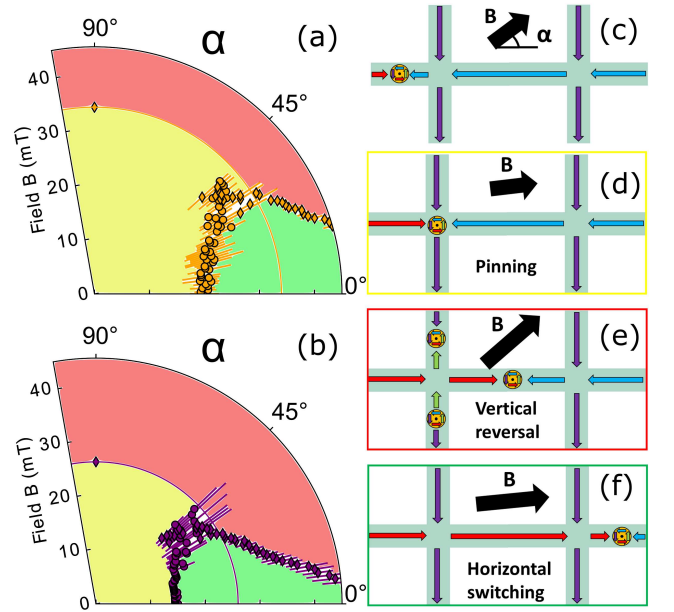}
\caption{(a) Experimental results of the measurement of cross n$^\circ$1 (orange diamonds and disks) under an externally applied field. The colours of the areas in (a) correspond to the events depicted in (d)-yellow, (e)-red, and (f)-green. (b) Similar plot as (a) for cross n$^\circ$2 (purple diamonds and disks). (c) Schematic representation of the starting magnetization configuration. The orange disk with the arrows represents a vortex DW. The colored arrows represent the magnetization direction. The black arrow indicates the applied field with strength B and angle $\alpha$. (d) and (e) Schematic representation of failure events, namely pinning of a DW in the cross and vertical arm reversal, respectively. (f) Schematic representation of the horizontal arm switching.}
\label{fig2}
\end{figure}

The cross-shaped intersection of magnetic nanowires is the key element of the closed-loop DW based multiturn-counter sensor device. The results of the measurement of magnetic switching of the branches of the cross under an applied field of strength B and angle $\alpha$ are presented in Fig. \ref{fig2} (a) and (b). To provide statistics, a system composed of 7 nominally identical crosses is measured for every geometry. At the beginning of all measurements, the system is initialized with a field magnitude of 70 mT oriented in the 225$^\circ$-direction. This procedure leaves the cross in a similar state as the one shown in Fig. \ref{fig2} (c), however, without the DW, i.e., the vertical arms magnetized downwards and the horizontal arm magnetized towards the left. 
In the experiment, we sense the reversing of the horizontal arms (Fig. \ref{fig2} (e) and (f)) by setting the MOKE sensitivity to a horizontal contrast. We then investigate a possible splitting of a DW (Fig. \ref{fig2} (e)) leading to the reversal of vertical arms by rotating the plane of incidence of the light by 90$^\circ$ yielding a sensitivity to a vertical contrast. In Fig. \ref{fig2} (a) and (b), the results of the two measurements are represented by disks and diamonds, respectively.

We interpret the data points as representing the boundaries between three characteristic processes. A schematic representation of these three possible outcomes is shown in Fig. \ref{fig2} (d, e, and f). The two first ones are failures for the operation of the device. The pinning of the DW at one of the crosses is the first one, where no change of the horizontal contrast of the right horizontal arms was observed in the microscope (b). The second one is a switching of one or more of the vertical branches that we term here vertical arm reversal. This splitting of the DW as indicated in Fig. \ref{fig2} (e) shows that one DW moves up or down a vertical arm thereby reversing it. Finally, the switching of the horizontal arm without a switching of the vertical arms shown in (d) is the desired behavior required for the functioning of the device concept.

Two different cross geometries (see Fig. \ref{fig1} (a) and (b)) are simultaneously measured and are plotted in Fig. \ref{fig2} (a) and (b). The differently coloured observable zones are each representing one of the three characteristic processes (red for vertical-arm reversal, yellow for pinning in the cross, and green for horizontal reversal solely). For cross n$^\circ$1, the center diagonal has the dimension 210$\sqrt{2}$ nm, while for cross n$^\circ$2, the diagonal is 300$\sqrt{2}$ nm. We find that qualitatively the same trend is followed for both geometries. 

At large angles, the distinction between the cross geometries is more obviously visible. This variation is attributed to the different nucleation fields inherent to the different cross diagonal lengths. In simple words, the narrower the diagonal, the larger the nucleation field due to increased shape anisotropy. 
Afterwards, measurements were performed with an angle $\alpha = 90^\circ$ to prevent horizontal DW motion. The obtained field strengths yielding a switching of the vertical arm mark a delimitation in Fig. \ref{fig2} (a) and (b) (orange and purple circular lines for cross n$^\circ$1 and 2, respectively). From this result, we conclude that for vertical-arm-reversal field values lower than the limit, the splitting of the DW leading to vertical-arm reversal is aided by the presence of a DW. The pinning of the DW in the center of the cross is represented by the yellow area. The depinning field is larger for cross n$^\circ$1 as compared to n$^\circ$2 due to the larger change in width while reaching the center of the cross. Finally, the green region represents the horizontal propagation necessary for the functioning of the device concept. The DW should reach the center of the cross solely for the field configurations (strength and angle) represented by this region. 

\begin{figure}[ht]
\centering
\includegraphics[scale=0.35]{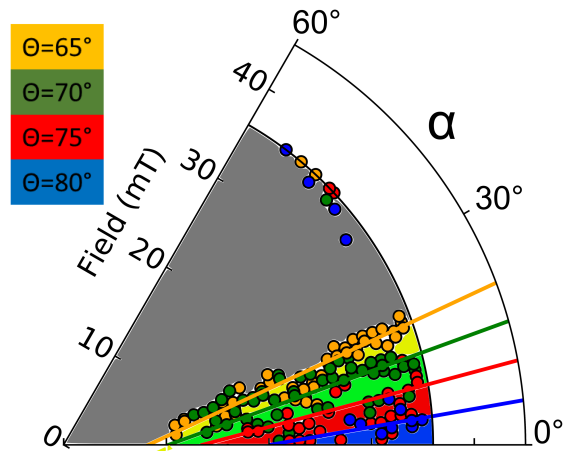}
\caption{Polar plot of the angular dependence for different angles $\theta$ of the syphon arm under an applied field B with angle $\alpha$. For $\theta$ = 65$^\circ$ (yellow, green, red and blue is propagation while gray is pinning), for $\theta$ = 70$^\circ$ (green, red and blue is propagation while gray and yellow is pinning), for $\theta$ = 75$^\circ$ (red and blue is propagation while gray, yellow, and green is pinning), and for $\theta$ = 80$^\circ$ (blue is propagation while gray, yellow, green and red is pinning). }
\label{fig3}
\end{figure}

The key problem of the closed-loop sensor concept is that a rotating field leads to the DW reaching the cross for values of $\alpha$ that lead to a vertical arm reversal. To overcome this problem and make sure that the DW only arrives at the cross for field angles that allow for a reliable horizontal propagation through the cross, we introduce a syphon element (see Fig. \ref{fig1} (c and e)). This geometrical element is designed to block the propagation of a DW for particular field directions. We next present the measurement of the depinning/propagation fields of a DW in several syphon geometries for various field configurations (strengths and angles). The syphon structures are always initialized before every measurement with the application of a 70 mT field oriented along 180$^\circ$-direction (see Fig. \ref{fig1} (c)).
The measurement is conducted as follows: the applied field strength was chosen, and the field was positioned at an angle of 45$^\circ$, placing a head-to-head vortex DW at the position shown by the DW in Fig. \ref{fig1} (c). The field was then rotated toward the horizontal direction (toward 0$^\circ$) by reducing the angle in steps of 1$^\circ$ until the switching of the magnetization. After that the field is increased by 1 mT, and the experiment is repeated. In Fig. \ref{fig3}, the measured values of 3 samples with different syphon angles are plotted together. The points separated from the rest of the distribution (i.e., around 45$^\circ$) represent a nucleation in the wire. The equation $H_{ext} = \frac{H_{p}}{|sin(\frac{3\pi}{2}-\theta-\alpha)|}$ with $H_p = 3.5$ mT describes the depinning of a DW in a wire submitted to a transverse and longitudinal field \cite{DH16}. The equation is plotted as a full line in Fig. \ref{fig3} and the experimental depinning values fit the theoretical description. The points are scattered in the vicinity of the line due to the stochasticity of the depinning process inherent to thermal activations and edge roughness present at the side of the wire. Interestingly, the edge roughness does not significantly increase the pinning value $H_p$ used in the equation, which is originally defined as a pinning field due to the curvature of the corners in a device \cite{DH16}. The impact of the edge roughness is limited to the absence of a depinning point for a particular field strength value. 

For the syphon angle $\theta$ = 80$^\circ$, some field strengths do not yield a depinning within the field range probed. We thus find a loss of reliability as the syphon angle is increased while the syphon operation is found to be reliable for angles of $\theta$ = 75$^\circ$ (red) and less. The nucleation field value for all the structures is identified to be 36 mT (black circular line), which defines the absolute maximum value usable for the whole structure of the device and is governed by the shape anisotropy.

\begin{figure}[ht]
\centering
\includegraphics[scale=0.6]{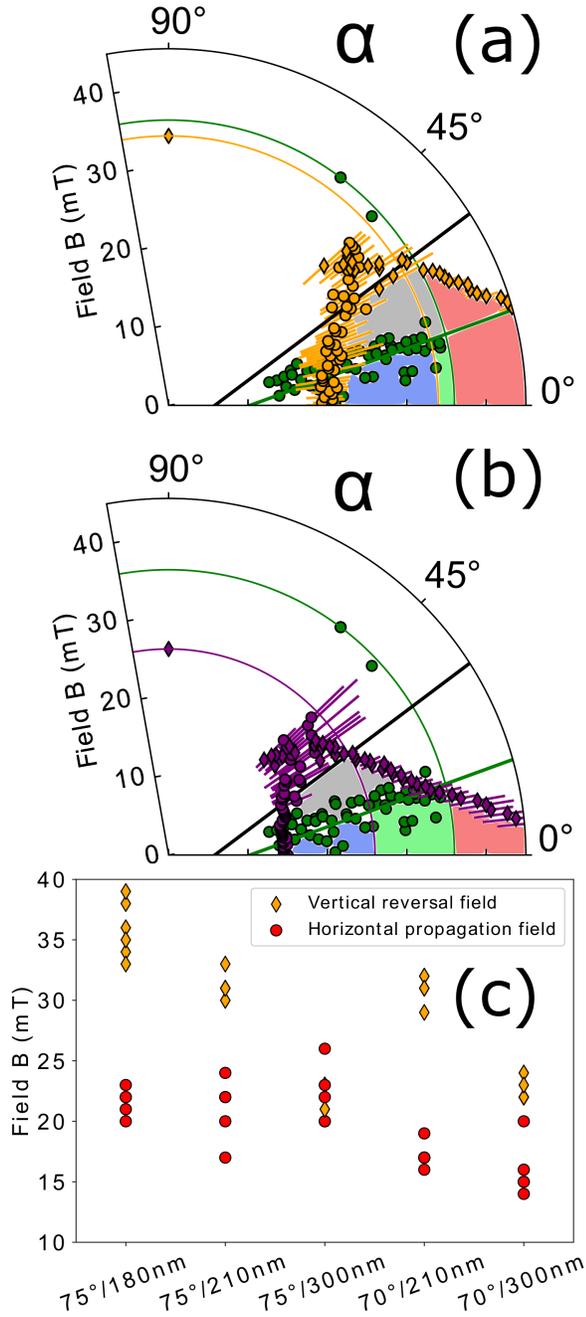}
\caption{(a) Polar plot of an expected complete device n$^\circ$1 comprised of cross n$^\circ$1 and a syphon ($\theta$ = 70$^\circ$). (b) Similar plot as (a) for device n$^\circ$2. (c) Plot on the measurement results of the depinning field (red circles) and the vertical-arm-reversal field (orange diamonds) in a complete structure with varying characteristics referenced in abscissa.}
\label{fig4}
\end{figure}

Finally we measure complete devices and compare the behaviour to what can be expected from the individual behaviour of the cross and syphon.
A device n$^\circ$1 is defined as containing cross n$^\circ$1, similarly for n$^\circ$2. In a combination plot (Fig. \ref{fig4} (a) and (b)), we overlap the angular dependence of a syphon with $\theta$ = 70$^\circ$ (Fig. \ref{fig3}) to the angular dependence of cross n$^\circ$1 and 2 (Fig. \ref{fig2} (a) and (b)), we call it the combined device as compared to the complete device, which is represented in Fig. \ref{fig1} (f). In the original concept of the closed-loop sensor a buffer region (details in Ref. \cite{Borsyphon17}) is defined as field configuration yielding a DW that could propagate in the cross without failure but is pinned in the syphon  (in gray in Fig. \ref{fig4} (a) and (b)). 
Combined device n$^\circ$1 exhibits a buffer region up to the nucleation field value for the syphon element while combined device n$^\circ$2 is limited to the meeting point between the syphon boundary (green line) and the vertical arm reversal (purple diamonds) at 34 mT in Fig. \ref{fig4} (b).

According to the original concept, the operating fields for the combined devices should be the field values at lower angles than the boundary formed by the syphon, and still exhibiting a buffer zone at larger angles, thus yielding the blue and green area in Fig. \ref{fig4} (a) and (b).  The red area is now representing the failure of the complete device concept due to a nucleation event occurring somewhere not only in the cross. This limit is set by the nucleation field of the syphon (see Fig. \ref{fig3}).

Looking at Fig. \ref{fig4} (c), the operating fields located in between the depinning and vertical-arm-reversal fields of complete devices (Fig. \ref{fig1} (f)) are represented. Interestingly, the vertical-arm-reversal fields are significantly reduced compared to the combined devices. This result is rather unexpected, and especially for device n$^\circ$2, the discrepancy between the combined device (\ref{fig4} (b)) and the complete one (\ref{fig4} (c)) is as high as 10 mT. 

For a complete device, the presence of the syphon limits the propagation of the DW to a narrow angular range of the applied field where the vertical component is small. The presence of the buffer zone makes the vertical arm reversal unlikely in this angular range. Thus if vertical arm reversal occurs, a DW cannot be present in the cross. The latter indicates that the limiting mechanism is the nucleation of new DWs in the center of the cross when the field is applied along one of its axes. The obtained field values for the vertical reversal of cross n$^\circ$1 and 2 without DW ($\alpha$ = 90$^\circ$) were used to create a demarcation (orange and purple circles) in Fig. \ref{fig4} (a) and (b), respectively. This demarcation separates vertical arm reversal with the aid of a DW at lower field values and reversal/nucleation at larger fields that would occur even without a DW.  It is of interest to note that these limits are corresponding to the vertical-arm-reversal values shown in Fig. \ref{fig4} (c). The reviewed field operating window for the combined device n$^\circ$1 in (a) is now from 20 mT to 34 mT (blue area), which corresponds to the ones of complete device (70$^\circ$/210 nm) in (c) exhibiting operating fields from 17 mT to 33 mT. For combined device n$^\circ$2, it is now from 15 mT to 26 mT (blue area) in (b) thus coinciding well with the complete device 70$^\circ$/300 nm in (c) that exhibits operating fields from 15 mT to 24 mT. Neither devices reach the absolute maximum nucleation value (36 mT) set by the cross-sectional dimensions of the syphon structure. The latter can as well be seen for other complete devices (75$^\circ$/210 nm, 75$^\circ$/300 nm, 70$^\circ$/210 nm, 70$^\circ$/300 nm) in (c) where the vertical arm reversal is setting the maximum limit and not the nucleation in the syphon.

Concerning the depinning values, we observe an increase for larger syphon angles in (c) resulting in the loss of the operating window for the device 75$^\circ$/300 nm while an operating window of 2 to 3 mT is still present for the device 70$^\circ$/300 nm. Similarly, a larger operating window is obtained for 70$^\circ$/210 nm as compared with 75$^\circ$/210 nm, due to equal vertical-arm-reversal fields set by the cross dimensions, and a smaller depinning field for the shallower syphon angle.

Theoretically the lower limit of the syphon angle for reliable operation sets 45$^\circ$ since lower angle values would allow domain walls to arrive at the cross for applied field angles ($\alpha$) that would yield a vertical arm reversal (e.g., if $\theta$ = 30$^\circ$ then an applied field at angle $\alpha$ = 55$^\circ$ would allow the propagation to the center of the cross but due to the larger y-component of the field, it would trigger the vertical arm reversal.)

From Fig. \ref{fig4} , we see that the nucleation in the center of the cross is the maximum operating applied field strength. The meeting point between the nucleation field values in the center of the cross (circular lines in Fig. \ref{fig4}  (a) and (b)) and the vertical arm reversal data (diamonds in Fig. \ref{fig4}  (a) and (b)) determines the experimentally-obtained maximum angle before a vertical arm reversal occurs. If we use the previously explained model for the syphon, we find a lower syphon angle limit for $\theta$ = 55$^\circ$ -60$^\circ$ represented in black in Fig. \ref{fig4} (a) and (b).

Finally, the device with the largest field operating window is 75$^\circ$/180 nm, which shows the highest value for the vertical-arm-reversal field likely limited by a nucleation event occurring in the syphon.
To increase this field operating window, a decrease of the width of the whole structure is necessary. 
Furthermore, since the center of the cross controls the upper limit of the operating window, very shallow angles of the syphon element (e.g., $\theta$ = 55$^\circ$) can be used thus also reducing the depinning field in the device.

We identify as the ultimate operation window the scenario where the depinning field is dominated by the center of the cross dimensions and not the syphon, and the nucleation field  is dominated by the cross-section of the wires in the device (e.g., syphon) and not the dimension of the cross, which provides clear guidelines for gauging the optimal performance that can be reached by this multiturn sensor device concept. 

To summarize, we present the experimental realization of the components of a closed-loop multi-turn sensor device, and we analyze the key components, which are a syphon together with a cross-shaped intersection architecture allowing for the reliable control of a DW under a rotating applied field. We observe a consistent behavior for different cross dimensions. The depinning field increases slightly with the decrease of the center diagonal dimensions, and the maximum nucleation field is largely enhanced providing a good indication that the reduction of the center is the key parameter thatcan be used to increase the operating window. The syphon element governs the field angles for which the DW arrives at the cross thus constituting a key element necessary for a reliable operation. It sets the maximum operating field for the complete device, which is identified to be the nucleation field in the syphon structure. Finally, the combination of the two elements reveals an unexpected limit set by the reversal of the cross center when a DW is not present, showing that the full device performance is not simply governed by the superposition of the individual performance parameters of the syphon and cross. This highlights that one needs to study the full device to obtain the correct operating conditions. We thus identify the central geometrical dimensions that set the limitation of the maximum field value of the field operating window up to the nucleation field of the syphon structure. As guidelines to obtain a large operating window one needs to choose a syphon angle sufficiently steep to allow the reliability of the interaction with the cross center ($\theta = 55^\circ$ or larger), and a center cross dimension small enough to reach the absolute maximum nucleation field set by the whole structure.

\section*{Acknowledgements}

The authors would like to acknowledge the WALL project for financial support. The work and results reported in this publication were obtained with research funding from the European Community under the Seventh Framework Programme - The people Programme, Multi-ITN “WALL” Contract Number Grant agreement no.: 608031, and a European Research Council Proof-of-Concept grant (MultiRev ERC-2014-PoC (665672)) as well as the German research foundation (SFB TRR173 Spin+X).


%

\end{document}